\documentclass{article}
\usepackage{arxiv}
\usepackage{graphicx}
\usepackage{mathptmx}
\usepackage[subrefformat=parens,labelformat=parens]{subcaption}
\usepackage{authblk}
\usepackage{amsmath}
\usepackage{paralist}
\usepackage{amsfonts}
\usepackage{amssymb}
\usepackage{amsbsy}
\usepackage{amsthm}
\usepackage{comment}
\usepackage{amsmath}

\usepackage[ruled, vlined]{algorithm2e}
\usepackage{bbm}
\usepackage{wrapfig}
\usepackage{booktabs}
\usepackage{multirow}
\usepackage{wrapfig,lipsum,booktabs}
\usepackage{hyperref}
\DeclareUnicodeCharacter{0301}{\'{e}} 

\begin{document}


\title{\Large 
From Discovery to Production: Challenges and Novel Methodologies for Next Generation Biomanufacturing
}
\author[  ~,1]{Wei Xie\thanks{Corresponding authors: w.xie@northeastern.edu}}
\author[  ~,2]{Giulia Pedrielli \thanks{Corresponding authors: giulia.pedrielli@asu.edu}}
\affil[1]{Northeastern University, Boston, MA 02115}
\affil[2]{Arizona State University, Tempe, AZ 85281}

\maketitle

\begin{abstract}
The increasingly pressing demand of novel drugs (e.g., gene therapies for personalized cancer care, ever evolving vaccines) with unprecedented levels of personalization, has put a remarkable pressure on the traditionally long time required by the pharma R\&D and manufacturing to go from design to production of new products. The revolution has already brought important changes in the technologies used within the industry. In fact, practitioners are increasingly moving away from the classical paradigm of large-scale batch production to continuous biomanufacturing with flexible and modular design, which is further supported by the recent technology advance in single-use equipment. In contrast to long design processes, low product variability (one-fits-all), and highly rigid systems, modern pharma players are answering the question: \textit{can we bring design and process control up to the speed that novel production technologies give us to quickly set up a flexible production run?} 

In this tutorial, we present key challenges and potential solutions from the world of operations research that can support answering such question. We first present technical challenges and novel methods for the design of next generation drugs, followed by the process modeling and control approaches to successfully and efficiently manufacture them.
\end{abstract}

\vspace{0.1in}


\section{INTRODUCTION}
\label{sec:intro}


{The rapidly growing biomanufacturing industry plays a significant role in supporting economy and ensuring public health.}
It has developed various innovative treatments for cancer, adult blindness, and COVID-19 among many other diseases. 
The industry generated more than \$300 billion revenue in 2019 with about 12\% annual growth rate \cite{Global2020}, and more than 40\% of the products in the development pipeline were biopharmaceuticals. 
However, drug shortages have occurred at unprecedented rates, especially in the COVID-19 pandemic.
It typically takes many years to discover a new bio-drug and develop the optimal production and delivery processes. The current manufacturing systems are unable to rapidly produce new drugs when needed when there is major public health issue. 

\textit{Biopharmaceutical manufacturing and delivery processes face critical {challenges}, including high complexity, high variability, lengthy lead time, and limited process observations.}
\begin{enumerate}
\item[(1)] \textbf{Long discovery processes and lack of connection with manufacturing.} Drug discovery is a highly experimental, expensive, complex and long process. Labs take up to a year to generate a potential design for few alternative versions of a drug (5$\sim$ 6). Algorithms can help in improving the quality and reducing the time to discovery, but substantial requirements are introduced in terms of data collection (e.g., collect information on failed experiments) to effectively adopt traditional data-driven optimization techniques. 
\item[(2)] \textbf{The discovery process is separated from the production. }The process of drug design and the experiments are usually performed with processes that are completely separated from the actual manufacturing pipelines used later on. This is justified in traditional bioproductions where manufacturing systems are highly rigid, expensive and can only be used for large volume productions. But novel systems (e.g., continuous manufacturing, single use technologies) allow for low volume productions, so we could \textit{move from experiments-based drug discovery to production-driven drug discovery}. This would lower the ``failure at manufacturing'', i.e., the scenario in which a drug passes the experimental phase but results impossible/very hard to manufacture and it is therefore rejected, thus triggering a new discovery process with consequent delays and costs.
\item[(3)] \textbf{High complexity and high uncertainty.}
Biomanufacturing process consists of numerous unit operations (such as fermentation, purification, formulation, and delivery). 
Biotherapeutics are produced in cells (or other living organisms, such as bacteria and yeast) whose biological processes are complex, and highly variable outputs depend on complex dynamic interactions of many factors. The upstream fermentation can impact on downstream purification cost and productivity.
New biotherapeutics (e.g., cell, gene, RNA, protein, and peptide therapies) require more advanced manufacturing protocols. 
For example, aspirin, a classical small molecule medicine is comprised of 21 atoms, whereas many of the antibodies (mAbs) protein drug substances are comprised of greater than 25,000 atoms. 
The drug size is correlated to the structural complexity of biopharmaceuticals. The molecular drug structure affects its function. 
In addition, the dynamic interactions of hundreds of factors can impact drug quality, yield, and production cycle time. 
The target protein and RNA can further degrade and have modifications during manufacturing and delivery
processes. \textit{Thus, the bioprocess is the product.} 
\item[(4)] \textbf{Very limited process observations.} The analytical testing time required by biopharmaceuticals of complex molecular structure is lengthy. 
Also, significant changes in the manufacturing process, such as new facilities, equipment, and raw materials, will typically trigger new regulatory requirements and clinical trials.
\end{enumerate}

Human error is frequent in biomanufacturing, accounting for 80\% of deviations~\cite{Ci15}. It is increasingly realized by the biopharma industry that optimization, machine learning, and simulation approaches, that can incorporate physics-based and experiments supported knowledge, are a key to the next generation of products and processes. This is because stochastic system modeling, analytics, and optimization methodologies can accelerate integrated and intensified manufacturing process automation, quality-by-design (QbD), and reduce human error, thus projecting biomanufacturing in Industry 4.0.

 \paragraph{Connecting drug discovery with manufacturing.}
Discovery and design of biological drugs and the associated, modular, production processes has the potential to dramatically reduce the currently prohibitive lead time from discovery to process design and manufacturing, while also increasing the quality, and reliability, of the final product. 

This integration is needed more than ever. Major technological developments are already enabling future bioproductions of large quantities of small volume (highly personalized) products. One of these advancements are ``single use technologies'', which encompass a range of products and technologies such as single-use disposable connectors, vessels, mixers, etc, which in turn enable fully automated and enclosed processes. They can support flexible and efficient manufacturing at scale and on-demand through 
 reducing (1) sterilization and cleaning costs, (2) contamination incidents, (3) storage needs, and (4) process downtime; see~\cite{Sandle2018epr}.
Therefore, single use technologies have the potential to impact existing medium to large volume biomanufacturing processes by enabling flexible manufacturing with modular design while reducing costs. The demand for such flexibility and variability in production batches (from few liters to tens of thousands) is already testing the capacity of pharmaceutical Contract Manufacturer Organizations (CMO).
For example, several research labs and startups of varying size are seeking the manufacturing capacity to produce vaccines for trials~\cite{Blankenship:2020}. %
In addition, single use technologies enable for the first time practical small volume bio-productions for personalized therapies, e.g., for cancer treatments~\cite{mock2016automated,zhu2017closed}.

\textbf{Limitations of state-of-art OR/OM approaches.}
Operation Research and Operations Management (OR/OM) methodologies can facilitate drug discovery, biomanufacturing system design and analysis, simulation model calibration \cite{lee2019review,arendt2012quantification,wang2019bayesian}
, sensitivity and uncertainty analyses \cite{saltelli2008global,baroni2014general}. 
 Mathematical programming methods \cite{lakhdar2007multiobjective,leachman2014automated}, 
supply chain planning \cite{fleischhacker2011planning}, Markov decision process and stochastic optimization \cite{
 kulkarni2015modular,martagan2018performance} 
 are developed to guide bioprocess decision making and optimization. \textit{However, existing OR/OM methodologies on process modeling, analytics, and optimization are general and they often fail to consider the physics underlying the specifics of drugs as well as the mechanisms activated during bioprocesses. These limitations hinder the application of these methodologies.}


In this paper, we first 
review the challenges and opportunities for novel algorithms in design and discovery of new biological drugs (e.g., mRNA vaccines, protein therapies). 
Section~\ref{subsec:RNA}
discusses the 
challenges and methodologies for molecular structure (folding) and functionality prediction with focus on Ribonucleic acids (RNAs), proteins, and nanoparticle delivery systems. 
Subsequently, Section~\ref{sec:HYBRID} presents a probabilistic knowledge graph (KG) hybrid modeling framework that can leverage the information from existing mechanistic models and facilitate learning from real-world data. Built on the hybrid model characterizing the risk- and science-based understanding on  bioprocessing mechanisms, we present the risk, sensitivity, and predictive analyses to support interpretable and robust decision making. Finally, we describe the control framework that can leverage these models. The presented reinforcement learning approach accounts for both inherent stochasticity and model uncertainty, to facilitate process development and control in Section~\ref{sec:RL}. To close the tutorial, Section~\ref{sec:summary} summarizes the key challenges in the biopharmaceutical industry from discovery to production and discuss the key OR methods and tools that are needed to be developed.




\section{THE MECHANICS OF DESIGN MOLECULAR FOLDING AND NANO-PARTICLES}
\label{subsec:RNA}

\begin{wrapfigure}{ht!}{0.4\textwidth}
	\centering
	\includegraphics[width=0.4\textwidth]{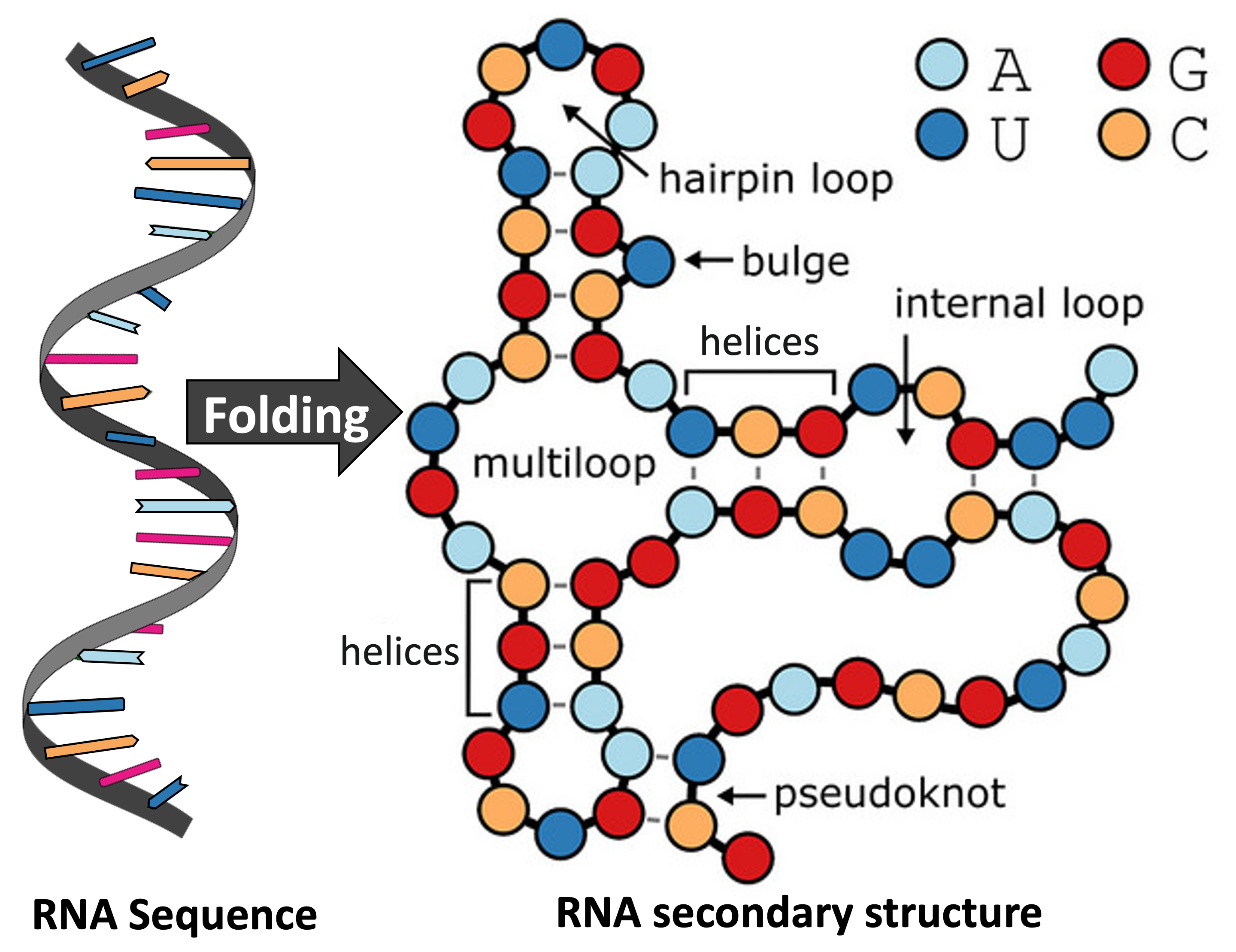}
	\vspace{-0.28in}
	\caption{
	\textit{RNA chain folding motifs.}
	}
	\label{fig:RNA-folding}		\vspace{-0.3in}
\end{wrapfigure}

Section~\ref{sec::strpred} focuses on the prediction of the folding configurations. We present methods for both the secondary (2D) and tertiary (3D) RNA structure, 
which directly impacts on RNA drug delivery and function. We also briefly review the structure prediction for proteins associated with dominant bio-drugs in the current biopharmaceutial industry and market. 
In Section~\ref{subsec:nanoparticles}, we investigate predicting the stability of peptides and nano-particles.
Nano-particle formulations have rapidly emerged as carriers in nucleic-acid therapies (i.e., DNA and RNA) 
to increase cellular uptake, support RNA delivery, and improve drug stability.




\subsection{Algorithms for Structure Prediction}\label{sec::strpred}

Ribonucleic acid (RNA) is a fundamental biological macromolecule, essential to all living organisms, performing a versatile array of cellular tasks including information transfer, enzymatic function, sensing, regulation, and structural function.
RNA has recently emerged as a promising drug target, with new therapeutic approaches aiming to develop drugs that target RNA rather than proteins. Moreover, designed RNA molecules are used in rapidly growing fields of synthetic biology and RNA nanotechnology, with applications to diagnostics, immunotherapy, drug delivery and realization of logical operations inside cells; see for example \cite{green2014toehold,geary2014single,han2017single}. 
In 
\cite{liu2022expertrna}, we propose a new framework, ExpertRNA, for the automatic folding of non-pseudoknotted secondary structures for RNA molecular compounds. ExpertRNA builds upon the fortified rollout algorithm and generalizes the architecture to allow for the consideration of multiple experts that can evaluate, at each iteration, the solutions generated by the base heuristic. 

\vspace{0.1in}

\textbf{RNA structure determines function.} 
Each RNA molecule is made up of a sequence of individual units, nucleotides (bases), which are of four common types, A, U, G and C  
(Figure~\ref{fig:RNA-folding}). 
Individual RNA sequences range in length from tens (tRNAs, siRNAs) to tens of thousands (viral genomes, long non-coding RNAs) and many contain further chemical modifications of the individual bases \cite{carell2012structure}. 
\textit{While identity is defined by sequence, the function of an RNA molecule is determined by its structure, i.e., the way nucleotides interact in space.}
Biochemists often break down {RNA structure} into four categories: \textbf{(1)} Primary structure refers to the sequence. \textbf{(2)} Secondary structure makes up the majority of the bonds in the structure  and includes the ``canonical base pairs'', where A pairs with U and G pairs with C, and the ``wobble base pair'', where G pairs with U.  This provides a 2D representation of the structure of the molecule and is the most commonly used level. \textbf{(3)} Tertiary structure defines 3D contacts via weaker, non-conical interactions. \textbf{(4)} Finally quaternary structure includes intermolecular interactions with other RNA molecules. Given the impact of structure on RNA functionality, the accurate computational prediction of the secondary and tertiary structure of RNA is an ongoing area of great interest in the computational biology community~\cite{calonaci2020machine,cruz2012rna}. 


 

\vspace{0.1in}

\textbf{{RNA secondary structure prediction. }}Most tools for secondary structure prediction~\cite{reuter2010rnastructure,hofacker2003vienna,zadeh2011nupack,zuker1981optimal} attempt to identify the structure that minimizes the free energy (FE) associated with the RNA molecule upon pairing a subset of the nucleotides, i.e., the energy released by folding a completely unfolded RNA sequence. The underlying assumption is that the structure with the lowest free energy is also the most likely structure the RNA will adopt. Equivalently, this family of approaches relies on the basic idea that 
the lower the FE the more stable the RNA structure will be. 
A \textit{first} challenge for this family of approaches is that it is not possible to exactly calculate the free energy due to the (i) incomplete understanding of the RNA molecular interactions, and (ii) the impractical computational cost of detailed kinetic simulation tools. As a result, several approximate models have been proposed in the literature~\cite{mathews1999expanded,xia1998thermodynamic,andronescu2010computational} 
to estimate the free energy associated with a given secondary structure. Most of the computational savings are a result of ignoring tertiary interactions. 
A \textit{second}, and possibly deeper, challenge is that this model assumes that an ``optimal'' structure is one that pairs nucleotides in a way that minimizes the free energy (MFE). However, RNA is known to fold cotranscriptionally~\cite{angela2018computationally}, i.e., the simultaneous transcription of two or more genes. Equivalently, RNA molecules might adopt a kinetically-preferred structure different from the global free energy minima.


In light of these challenges, alternative approaches to structure prediction have been proposed. \textit{Stochastic kinetic folding algorithms}~\cite{isambert2000modeling,sun2018predicting} approximate the folding kinetics of RNA molecules as they are transcribed. Data driven approaches have also started to become popular that use machine learning to evaluate structures rather than FE or kinetic models. These include ContraFold, DMfold, and structure prediction with neural networks~\cite{do2006contrafold,calonaci2020machine}. 
Furthermore, in the attempt to achieve advantages of model or data driven approaches, methods have been proposed that attempt to aggregate multiple information sources to get more accurate secondary structure prediction. Within the data driven category, some examples of information sources are the experimentally determined SHAPE data~\cite{lucks2011multiplexed,low2010shape}, and evolutionary covariation information~\cite{calonaci2020machine}. On the model driven side, 
statistical ensemble approaches are used to boost the solutions obtained by the different FE driven folding algorithms. To the knowledge of the authors, ensemble methods allow to mix solutions from different algorithms only upon completion, i.e., they do not enable interaction among the algorithms while they are running~\cite{aghaeepour2013ensemble}. 
A recent survey on a set of secondary structure prediction tools has reported mixed results, with data-driven approaches generally outperforming the ones based on nearest-neighbor free energy models, and with model-based ensemble approaches showing competitive results~\cite{wayment2020rna}. 

\vspace{0.1in}

\textbf{Tertiary Structure Prediction. } Concerning the tertiary structure prediction problem, fewer approaches can be found in the literature~\cite{miao2017rna,watkins2020farfar2}. 
In fact, the prediction of tertiary structures is particularly challenging, and most prediction methods only work for short RNA sequences (tens of nucleotides). 
Data driven approaches have attracted attention also for the tertiary structure prediction. However, their accuracy remains limited due to the small number of 3D RNA structure data sets available for model training and verification. 



\vspace{0.1in}

\begin{figure}
\centering
\begin{minipage}{.47\textwidth}
  \centering
  \vspace{-0.3in}  
  \includegraphics[width=1\linewidth]{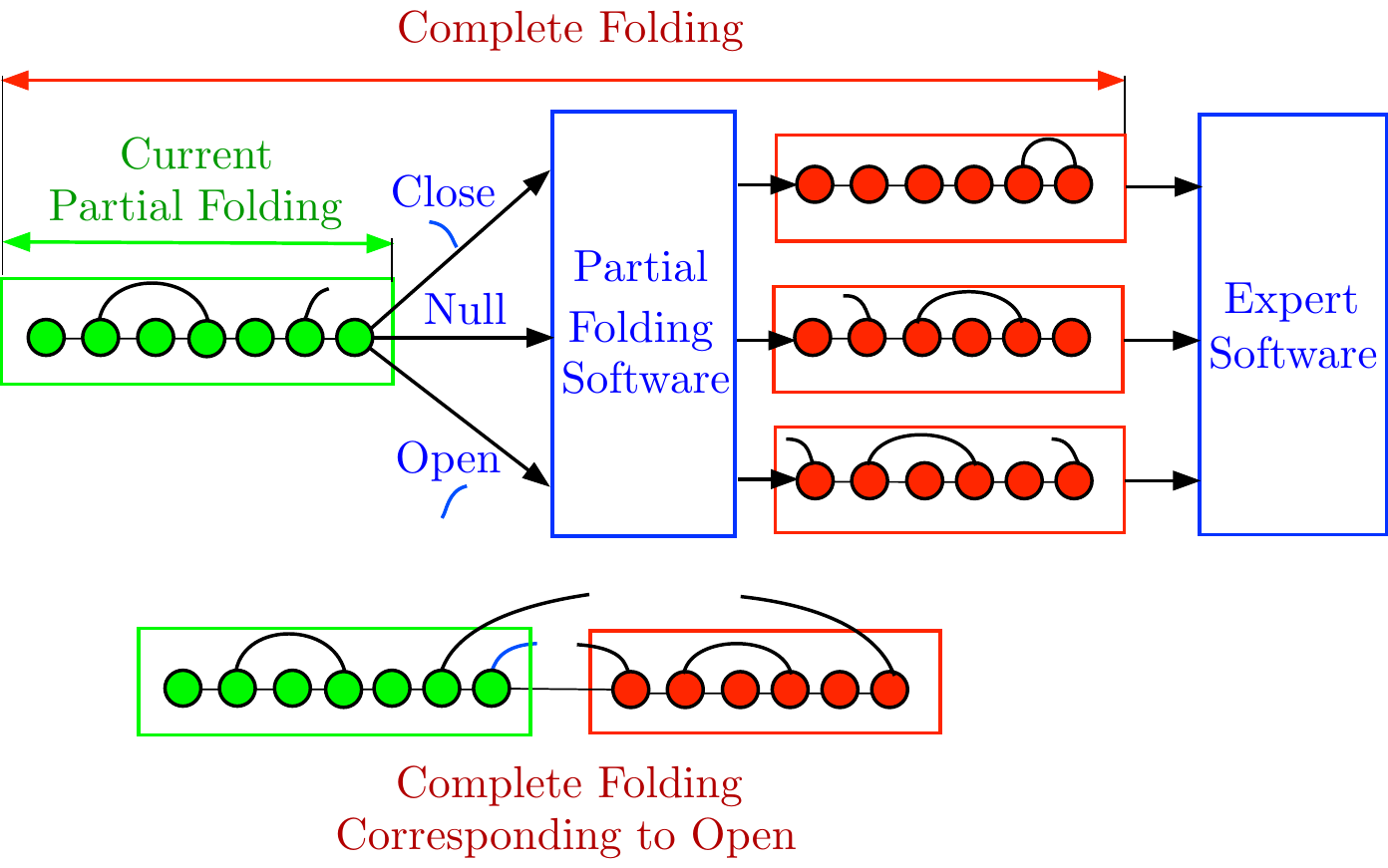}
  \captionof{figure}{Overview of the ExpertRNA algorithm (source in~\protect\cite{liu2022expertrna}).}
  \label{fig::XRNA}
\end{minipage}\hfill%
\begin{minipage}{.53\textwidth}
  \centering
  \includegraphics[width=1\linewidth]{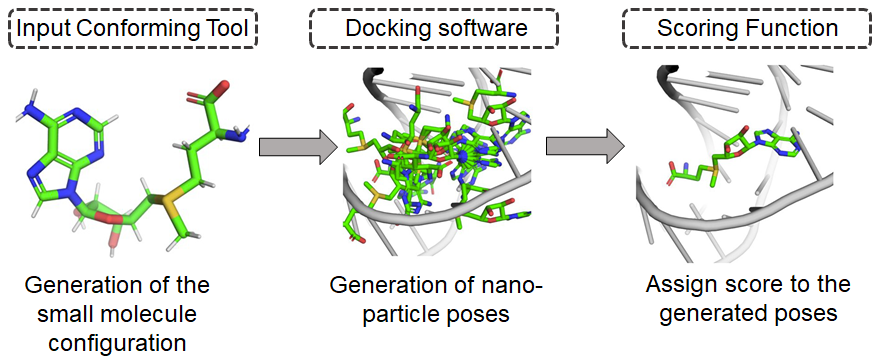}
  \captionof{figure}{
  Components for nano-particle docking.}
  \label{fig:my_label}
\end{minipage}
\vspace{-0.1in}
\end{figure}

\paragraph{The ExpertRNA framework for RNA structure prediction.}
Stemming from the observation that several folding algorithms 
have been proposed in the literature for secondary and, even if fewer, for tertiary structure prediction \cite{miao2017rna,watkins2020farfar2}, without any approach dominating the other, we propose the idea to build a framework that can exploit several folding tools and criteria to evaluate the quality of a folded sequence, {during} the algorithm execution \cite{liu2022expertrna}. 
The aim of our approach is to achieve a better RNA structure prediction quality.
\textit{Figure~\ref{fig::XRNA} shows the structure of our ExpertRNA approach with its two main algorithmic components: (i) the partial folding; and (ii) the expert software.} 
To mimic interatomic interactions, the algorithm sequentially adds elements to the incomplete structure (``current partial folding'' in Figure~\ref{fig::XRNA}), which we initialize to be the empty set. The first nucleotide is chosen as the first element of the input sequence provided by the user.
At each step, the subsequent nucleotide is selected, and we can choose whether to simply \textit{sequence} it to the last assigned nucleotide (``Null'' action in Figure~\ref{fig::XRNA}) or pair it with any nucleotide in the existing structure (``Close'' in Figure~\ref{fig::XRNA}), or pair it with an element still to be assigned (``Open'' in Figure~\ref{fig::XRNA}). The definition of these actions is motivated by the physical laws that govern molecular bonding (as previously specified in feasibility determination).

\vspace{0.1in}

\paragraph{Algorithms for protein structure prediction}\label{sec::protein}
Several approaches have been proposed in the literature for protein structure prediction. Also in the case of proteins, we distinguish primary, secondary, and tertiary structure prediction. It is important to predict protein structure due to the implications in medicine as well as biotechnnology. Several algorithms have been proposed in the area with an increasing push into deep learning mainly justified by the exhaustive data sets freely available for proteins. 
Given different folding software to allow constraints to be provided by our method,
we investigate how to embed existing tools such as PEP-FOLD and variants~\cite{thevenet2012pep,shen2014improved,lamiable2016pep}, AWSEM~\cite{jin2020awsem}, Rosetta~\cite{chaudhury2010pyrosetta}, and compare to the Maestro software from Schrodinger LLC. Importantly, once every two years the Critical Assessment of protein Structure Prediction (CASP) experiments are held to assess the state of the art in the field in a blind fashion, by presenting predictor groups with protein sequences whose structures have been solved but have not yet been made publicly available. DeepMind’s entry, AlphaFold, placed first in the Free Modeling (FM) category, which assesses methods on their ability to predict novel protein folds (the Zhang group placed first in the Template-Based Modeling (TBM) category, which assess methods on predicting proteins whose folds are related to ones already in the Protein Data Bank)~\cite{alquraishi2019alphafold}. DeepMind’s success generated significant public interest. Their approach builds on two ideas developed in the academic community during the preceding decade: (i) the use of co-evolutionary analysis to map residue co-variation in protein sequence to physical contact in protein structure, and (ii) the application of deep neural networks to robustly identify patterns in protein sequence and co-evolutionary couplings and convert them into contact maps~\cite{marx2022method}.

\subsection{Predicting Stability of Nano-particles for RNA Formulation and Delivery}
\label{subsec:nanoparticles}

When nanoparticles as RNA delivery system are of interest, a foundational question arises related to the docking of multiple molecules; see the illustration in Figures~\ref{fig:my_label} and \ref{fig::mRNA-LNP}.
In case the molecules are of the same family (e.g., peptides with peptides, RNAs with RNAs), the problem can be brought back to the 
folding prediction approaches in Section~\ref{sec::strpred}. However, in the case of assembly of heterogeneous bodies (i.e., peptides with RNA), new challenges arise. In this case, we normally refer to the problem of \textit{docking} of molecules, a molecular modeling which can estimate the preferred orientation of one molecule to a second and further predict the type of signal produced and the strength of binding affinity between two molecules using scoring functions.
The challenges in docking are very different from those identified in folding. Molecular docking processes are typically composed of two steps and usually a large molecule and a small molecule are considered for binding~\cite{sellami2021virtual}: \begin{inparaenum} \item[(1)] The conformation of the small molecule (e.g., ligand, peptide, lipid) is predicted together with the orientation and position with respect to the binding site of the larger molecule (e.g., protein, DNA, RNA). Such location and conformation is commonly referred to as the \textit{pose} of the nano-particle. The quality of the pose requires assessment. Such evaluation can be performed by a scoring function. 
    Ideally, a scoring function should be capable to recover the experimental \textit{binding} (true) and rank it highest among all the solutions proposed by the sampling algorithm.
    \item[(2)] A second, but especially challenging task would be to attempt the scoring of active vs. inactive compounds. This task is rarely accomplished due to the interplay of several factors that are external to the nano-particle.
\end{inparaenum}

We focus on the search algorithms that have been designed to \textit{efficiently} predict the docking pose. The process of docking a target and a small molecule falls into the class of NP-hard problems due to the non countability of the number of possible poses. Hence, search becomes the approach to identify candidate solution and improving on those. In the literature, search methods can be classified into deterministic (also referred to as {systematic}) and stochastic~\cite{stanzione2021use}. \textit{Systematic search methods} sample within the binding molecule search space at predefined intervals and are deterministic. Within this class, we can still differentiate between exhaustive, fragmentation
or conformational ensemble methods. The main difference between them is in the approach they take to deal with the binding molecule flexibility. In exhaustive search methods, for example, the docking is performed by systematically rotating all possible rotatable bonds in the binding molecule at a given interval. The drawback of this family of approaches is computational in nature as the number of possible combinations to consider goes with the number of rotatable bonds in the the binding molecule. An common exhaustive sampling method is Glide presented in~\cite{friesner2004glide,halgren2004glide}. Fragmentation represents an attempt to improve on the computational efficiency by incrementally forming binding over fragments that the binding molecule is divided into. An approach that relies on fragmentation is FlexX~\cite{rarey1996fast}. Finally, in conformational ensemble methods, the binding molecule flexibility is represented by rigidly docking an ensemble of pre-generated conformations, thus improving the approach efficiency since using this approach removes the computational cost due to the exploration of the conformational space.

In \textit{stochastic algorithms}, the binding molecule orientations and conformations are sampled by making changes to the molecule that are informed by random score values iteratively generated by a random algorithm. The orientation, conformation change is then treated as a incumbent that is accepted or rejected according to an algorithm-dependent criterion. The advantage of stochastic algorithms is that they can generate large ensembles of molecular conformations and explore a broader range of the
energy landscape increasing the probability of finding a global energy minimum.
However, this comes at computational cost. Examples are genetic algorithm, Monte
Carlo, ant colony optimization (ACO) and tabu search methods. GOLD in~\cite{jones1997development} uses a genetic algorithm, DockVision~\cite{hart1992multiple} is an example of docking program using Monte Carlo stochastic method where the probability to accept a random change is calculated by using the Boltzmann probability function. An example of ACO-based approach is PLANTS~\cite{korb2006plants}, while PSI-DOCK uses a tabu search~\cite{pei2006psi}.

\vspace{0.1in}

\textbf{Data Sets.}
Search algorithms usually rely on data sets to retrieve potential components of the nano-particle to be assembled. In these datasets, the crystal structures of the complexes are specified using different microscopy technologies (with potentially different resolutions). Prior to any docking study, the virtual compounds database that will be screened must be carefully selected and prepared. This compound collection, often referred to as the \textit{virtual library}, can encompass up to millions of compounds. Already prepared virtual libraries can be used, but users can also generate their own. Commercial databases represent an important source of compounds for virtual screening (often containing more than 1 million molecules). Suppliers commonly offer free access to the files containing molecules structures in several formats (2D or 3D). These databases undergo frequent
updates, new products being added while other being either removed or out of stock. Virtual screening libraries constructed from these databases should ideally be prepared when the whole virtual screening protocol is settled and ready to be used.

A particular category of databases are \textit{bioactivity databases} providing knowledge on biological targets and their modulation mechanisms. These data are frequently used for data-set benchmarking in the context of docking protocols design prior to prospective virtual screenings and
to construct predictive models of activity~\cite{huang2006benchmarking,lagarde2015benchmarking,mysinger2012directory}. 
Examples of databases in this category are ZINC~\cite{sterling2015zinc} or PubChem~\cite{kim2019pubchem}.

Finally, \textit{natural products data bases} are available. Natural products were the first drugs ever used and have always been a source of drugs. In a recent retrospective study, it has been reported that, between January 1, 1981 and September 30, 2019, 23.5\% of all new approved drugs and 33.6\% of new approved small molecules drugs were natural products or derivatives of natural products~\cite{newman2020natural}. The chemical space covered by natural products is quite dissimilar to the one occupied by synthetic drug-like compounds~\cite{morrison2013natural} and natural products are believed to constitute promising starting point for drug discovery~\cite{rodrigues2016counting}. Hence, these data sets can represent a source for virtual screening libraries. Numerous databases of this type are available; please refer to~\cite{sorokina2020review} and~\cite{chen2017data}).

\vspace{0.1in}

\textbf{Scoring Functions.}
There are three important applications of scoring functions in molecular docking~\cite{huang2010scoring}. 
The \textit{first} of these is 
 the determination of the binding mode and site of small molecule to its target.
Specifically, once a target has been defined, molecular docking generates hundreds of thousands of possible binding orientations/conformations for the small molecule (e.g., ligand, peptide) at the active site around the target molecule. The scoring function is used to rank such small molecule orientations/conformations by evaluating the binding tightness of each of the candidate complexes. Ideally, the scoring function would rank the highest the experimentally determined binding mode. Given the determined binding mode, scientists would be able to gain a deep understanding of the molecular binding mechanism and to further design an efficient drug by modifying either the target or the small molecule. It is important to highlight that, instead of scoring functions, other computational methodologies based on molecular dynamics or Monte Carlo simulations could be used to model the dynamics of the binding process thus resulting in a more accurate prediction of binding affinity. However, these models result in computationally prohibitive free energy calculations, ultimately resulting impractical for the evaluation of large numbers of molecular complexes. As a result the application of high fidelity simulation techniques is generally reduced to predicting binding affinity in small nano-particles~\cite{ain2015machine}.

The \textit{second application} of a scoring function, related to the first, is the prediction of the \textit{absolute binding affinity} between the active compounds. This is particularly important in lead optimization, i.e., the process to improve the tightness of binding for low-affinity hits or lead compounds that have been identified. In this phase, an accurate scoring function can greatly increase the optimization efficiency and save costs by computationally predicting the binding affinities between the conformed small molecule and the target before the much more expensive synthesis and experimental steps.

The \textit{third application}, is related to the foundational task of structure-based design, that fundamentally attempts to identify the potential drug hits/leads for a given target by searching a large compound database, this is commonly referred to as \textit{virtual database screening}. 
A reliable scoring function should be able to associate higher rank to known binders following their binding scores during database screening. In fact, due to the expensive cost of experimental screening and sometimes unavailability of high-throughput assays, virtual database screening has played an increasingly important role in drug discovery.

Classical scoring functions can be classified into three groups: forcefield, knowledge-based and empirical~\cite{huang2006molecular,gohlke2000knowledge,krammer2005ligscore,guvench2009computational,ballester2010machine}. 
An alternative to the classical approach to the design of scoring functions, a non-parametric machine-learning approach can be taken to implicitly capture the binding interactions that are hard to explicitly model by classical approaches in a computationally efficient way. By not imposing a particular functional form for the function, the collective effect of intermolecular inter-actions in binding can be directly inferred from experimental data, which should lead to scoring mechanisms that are characterized by greater generality and, consequently, prediction accuracy.

\section{BIOMANUFACTURING HYBRID MODELING AND ANALYTICS}
\label{sec:HYBRID}


Driven by the critical challenges and needs of biopharmaceutical manufacturing, we propose the probabilistic knowledge graph (KG) hybrid model characterizing the spatial-temporal causal interdependencies of critical process parameters (CPPs) and critical quality attributes (CQAs) ~\cite{zheng2022opportunities,xie2022interpretable}. 
To faithfully represent underlying bioprocessing mechanisms, this hybrid model can capture the important properties, including 
nonlinear reactions, partially observed state, and nonstationary dynamics. 
It has time-varying kinetic coefficients (such as molecular reaction rates) with uncertainty representing batch-to-batch variation. 
To avoid the evaluation of intractable likelihood, we further investigate a computational Bayesian inference approach, called Approximate Bayesian Computation (ABC), to efficiently approximate the posterior distribution. 
Then, assisted by the Bayesian KG, accounting for inherent stochasticity and model uncertainty, we present \textit{interpretable} risk, sensitivity, and predictive analyses.
This study can support: (1) biomanufacturing stochastic decision process (SDP) mechanism online learning; (2) bioprocess soft sensor monitoring (such as tracking latent metabolic state associated with therapeutic cell function and product quality); and (3) optimal and robust process control.

\vspace{0.1in}
\textbf{Illustration Example: mRNA lipid nanoparticle formulation process.}
Here we use mRNA lipid nanoparticle (mRNA-LPN) formulation process as an illustration example; see Figure~\ref{fig::mRNA-LNP}. 
Lipid-based formulations have rapidly emerged as carriers in nucleic-acid therapies (i.e., DNA and RNA) 
to 
increase cellular uptake, support RNA delivery, and improve drug stability. The mRNA lipid nanoparticle (LNP) formulation utilizes microfluidic or T-junction mixing to rapidly combine an ethanol phase containing  hydrophobic lipids and an aqueous phase containing mRNA in a buffer. Then, the self-assembly of LNPs with mRNA is driven by the hydrophobic and electrostatic force field that is influenced by the design of lipids, the selection of LNP formulation and CPPs, as well as the phases of mRNA-LNP complex. The pH-responsive ionizable cationic lipids have the surface charge modulated, controlling the efficient binding with the oppositely charged polynucleotides, which will support self-assembly, influence mRNA-LNP thermodynamic stability, prolong the circulation time of mRNA-LNP complexes, facilitate endosomal escape and mRNA release, and increase their therapeutic.
Therefore, the dynamics and variations of mRNA-LNP formulation process output trajectory depends on complex interactions of (1) the design of lipids and (2) {CPPs} (e.g., flow rate ratio, total flow rate, temperature, lipid choices, buffer choices), which impacts on the quality of LNP delivery system specified {CQAs} including a) RNA integrity level; b) species composition/concentrations/phases, particle size distribution, zeta potential, surface charge; and c) mRNA-LNP, bound/unbound mRNA.


In Section~\ref{sec:operationUnits}, we present the macro-phases or operation units involved in the manufacturing of a bioproduct, while Sections~\ref{sec::KGbioprocess} and \ref{sec:RL} illustrate novel methods developed for the efficient modeling, analytics, and control of such process when targeting novel synthetic products (e.g., variations of the mRNA-LPN).


\subsection{Operation Unites
of a Biomanufacturing process}
\label{sec:operationUnits}

Biopharmaceutical manufacturing process is crucially important to determine product quality and productivity.
It typically includes the main unit operations listed below.
\begin{inparaenum}\item[(1)]\textbf{Fermentation and Drug Substance Synthesis}: Living organisms (e.g., cells, yeasts) are mixed with appropriate medium and enzymes under carefully controlled conditions to grow and 
synthesize the target drug substance. 
The byproducts or unwanted impurities are also generated at the meantime, which impacts downstream purification operation and cost. 
During different growth and production phases of cell and yeast life cycle, different media compositions and feeding strategies are used to improve the productivity and reduce the waste generation. 
\item[(2)] \textbf{Centrifugation(s)}: The centrifuge device is used for separation of particles, e.g., cells, subcellular organelles, viruses, large molecules such as proteins, from a solution according to their size, shape, density, viscosity of the medium and rotor speed, during which bulk of impurities would be removed. 
\item[(3)]\textbf{Chromatography and Purification}: As the mixture of solutes flows through a packed resin bed, specific solutes are separated as they are bound. Chromatography serves as the most critical part for purification, and it usually determines the purity of product. Since removing more impurities often results in  removing more drug substance during chromatography step, there is often a trade-off between productivity and purity.
\item[(4)] \textbf{Filtration}: It is applied at several stages for capture (i.e., concentrate the product), intermediate purification, and polishing (i.e., eliminates trace contaminants and impurities) purpose.
\item[(5)]\textbf{Formulation and Filling}: To maintain the safety and efficacy of the drug substance during the storage, delivery, and facilitate the patient absorption of active pharmaceutical ingredient (API), the purified drug substance is usually formulated with carefully selected excipients and nanoparticle carriers into stable drug products and filled into dose containers~\cite{patro2002protein}.
\item[(6)]\textbf{Freeze Drying}: It is used to stabilize bio-drugs through removing water or other solvents from the frozen matrix and converting the water directly from solid phase to vapor phase through sublimation. 
Freeze drying is critical for immobilizing the bio-drug product in storage and delivery, as the kinetics of most chemical and physical degradation reactions are significantly decreased. 
\item[(7)]\textbf{Quality Assurance/Control (QA/QC)}: QA and QC are performed to ensure the quality of the raw material selection, the production process, and the final bio-drug product.
\end{inparaenum}
In sum, Step (1) belongs to upstream fermentation and drug substance synthesis, Steps~(2)--(4) belong to downstream purification, and Steps~(5)--(7) are for finished drug filling/formulation, freeze drying, 
and product quality control testing.

There are interactions of hundreds of factors at different productions steps impacting  drug quality, yield and production cycle time. 
These factors can be divided into critical process parameters (CPPs) and critical quality attributes (CQAs) in general; see the definitions of CPPs/CQAs in ICH-Q8R2 \cite{guideline2009pharmaceutical}.
\begin{itemize}
	\item[CPP:] At each process unit operation, CPPs are defined as critical process parameters whose \textit{variability} impacts on product CQAs, and therefore should be monitored and controlled to ensure the process produces the desired quality.
	
	\item[CQA:] A physical, chemical, biological, or microbiological property that should be within an appropriate limit, range, or distribution to ensure the desired {product} quality.
\end{itemize}

\subsection{KG hybrid model for Biomanufacturing Stochastic Decision Process}
\label{sec::KGbioprocess} 
The probabilistic KG hybrid model proposed in  \cite{zheng2022opportunities} and \cite{xie2022interpretable} can provide the risk- and science-based understanding of  underlying {stochastic decision process (SDP) mechanisms} for {controlled} production processes. The input-output relationship in each step is modeled by \textit{a hybrid (``mechanistic+statistical") model} that can leverage the prior knowledge on biophysicochemical mechanisms from existing mechanistic models and further advance scientific learning from process data. 
Specifically, at any time $t$, the process state $\pmb{s}_t$ (such as CQAs) composed of observable and latent state variables, i.e., $\pmb{s}_t=(\pmb{x}_t,\pmb{z}_t)$ (e.g., particle size distribution, RNA sequence and integrity level), 
and CPPs action $\pmb{a}_t$ (e.g., temperature, mixing flow rate, pH) interactively influence on the dynamics and variations of output trajectories (e.g., mRNA-LNP formulation and self-assembling processes).
Given the existing nonlinear ODE/PDE-based mechanistic model (such as biomolecular dynamics, thermodynamics, molecular interactions of mRNA and lipids which can affect the RNA integrity), represented by  
${\mbox{d}\pmb{s}}/{\mbox{d}t} = \pmb{f}\left(\pmb{s},\pmb{a}; \pmb\beta\right),$
by applying the finite difference approximations on derivatives, 
we construct the hybrid model for state transition, 
\begin{eqnarray}
       \pmb{x}_{t+1} &=& \pmb{x}_t + \Delta t \cdot \pmb{f}_x(\pmb{x}_t,\pmb{z}_t,\pmb{a}_t; \pmb{\beta}_t) + \pmb{e}^{x}_{t+1},
       \label{equ:hybridobs} 
       \nonumber \\
    \pmb{z}_{t+1} &=& \pmb{z}_t + \Delta t \cdot \pmb{f}_z(\pmb{x}_t,\pmb{z}_t,\pmb{a}_t; \pmb{\beta}_t) + \pmb{e}^{z}_{t+1},
    \label{equ:hybridlatent} \nonumber 
\end{eqnarray}
with unknown random kinetic coefficients $\pmb{\beta}_t\in \mathbb{R}^{d_\beta}$ (e.g., particle clustering rates) accounting for batch-to-batch variation. The function structures of $\pmb{f}_x(\cdot)$ and $\pmb{f}_z(\cdot)$ are derived from $\pmb{f}(\cdot)$ in the mechanistic models.
The residual terms 
are modeled by  multivariate Gaussian distributions $\pmb{e}_{t+1}^{x} \sim \mathcal{N}(0,V^{x}_{t+1})$ and $\pmb{e}_{t+1}^{z} \sim \mathcal{N}(0,V^{z}_{t+1})$ with zero means and covariance matrices $V^{x}_{t+1}$ and $V^{z}_{t+1}$ by applying CLT.
The kinetic coefficients $\pmb\beta_t$ can change cross different phases of bioprocess accounting for the fact that the process dynamics can be time-varying. 
{The statistical residual terms $\pmb{e}_t=(\pmb{e}_t^x,\pmb{e}_t^z)$ allow us to account for the impact from bioprocess noises, raw material variations, 
ignored CPPs, sensor measurement errors, and other uncontrollable factors (e.g., contamination) occurring at any time step $t$.}

\begin{figure}
\centering
\begin{minipage}{.46\textwidth}
\vspace{0.03in}
  \centering
  \vspace{-0.0in}  
  \includegraphics[width=1\linewidth]{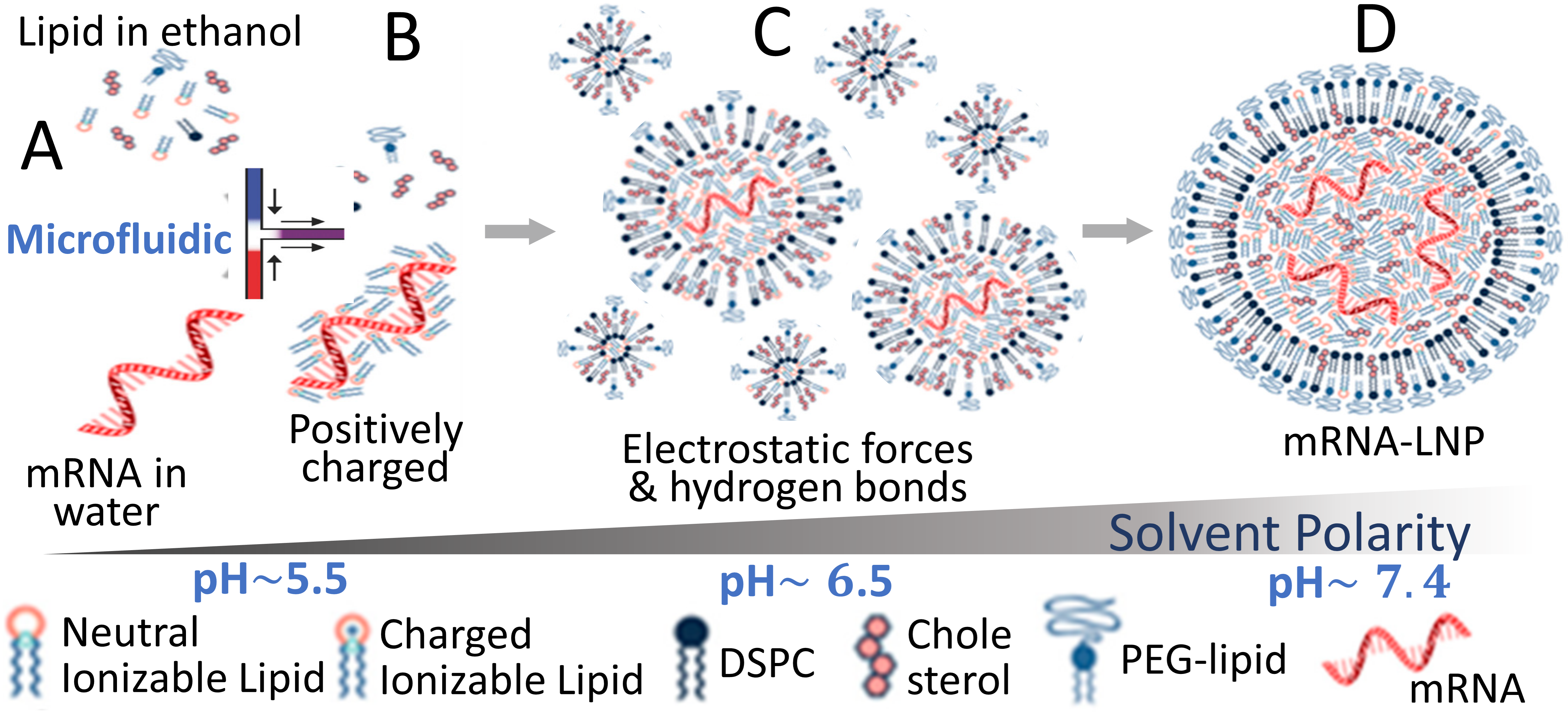}
  \vspace{0.in}
  \captionof{figure}{An illustration of mRNA lipid nanoparticle formulation and self-assembly process
  (figure adapted from \cite{buschmann2021nanomaterial}).}
  \label{fig::mRNA-LNP}
\end{minipage}\hfill%
\begin{minipage}{.54\textwidth}
\vspace{-0.2in} 
  \centering
  \includegraphics[width=1\linewidth]{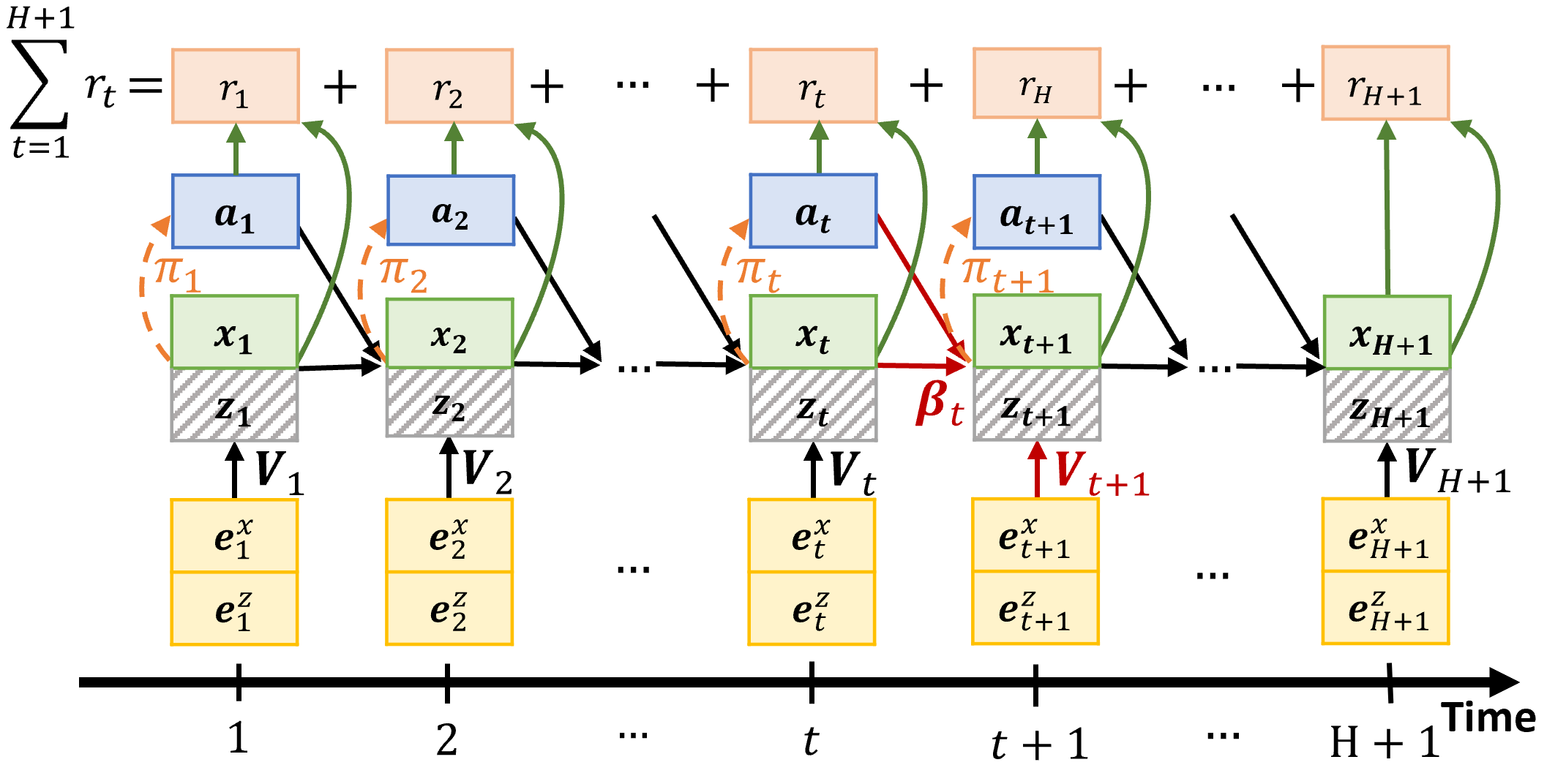}
  \captionof{figure}{Policy-augmented KG network for SDP.}
  \label{fig:hybrid_example}
\end{minipage}
\vspace{-0.1in}
\end{figure}

\begin{sloppypar}
The probabilistic KG
of integrated biomanufacturing process can be visualized by a directed network as shown in Figure~\ref{fig:hybrid_example}.
The observed state variables $\pmb{x}_t$ and the latent state variables $\pmb{z}_t$ are represented by solid and shaded nodes respectively. The directed edges represent causal interactions.
At any time period $t+1$, the process state output node
$\pmb{s}_{t+1}=(\pmb{x}_{t+1},\pmb{z}_{t+1})$ depends on its parent nodes: $\pmb{s}_{t+1}=\pmb{f}(Pa(\pmb{s}_{t+1});{\pmb\theta}_t)$
with $Pa(\pmb{s}_{t+1})=(\pmb{s}_t,\pmb{a}_t,\pmb{e}_{t+1})$ and model parameters denoted by $\pmb{\theta}_t$. 
To represent the underlying controlled SDP, we create a \textit{policy augmented KG network} by including additional edges: 1) connecting state $\pmb{s}_t$ to action $\pmb{a}_t$ representing the causal effect of the policy, $\pmb{a}_t=\pi_t (\pmb{s}_t|\pmb{\phi})$ specified by parameters $\pmb{\phi}$; and 2) connecting actions and states to the immediate reward $r_t(\pmb{s}_t,\pmb{a}_t)$  (e.g., cost and RNA production).
\textit{This KG network models how the effect of current state and action, $\{\pmb{s}_t,\pmb{a}_t\}$, propagates through mechanism pathways impacting on the output trajectory and the accumulated reward.} 

\end{sloppypar}


\vspace{0.1in}
\begin{sloppypar}
\textbf{Bayesian KG representing risk- and science-based mechanism understanding.} Leveraging the information from existing mechanistic models and heterogeneous online/offline measurements, the probabilistic KG hybrid model represents the understanding of bioprocess mechanisms. It allows us to inference the latent state (e.g., RNA and nanoparticle binding strength, mRNA-LNP folding structure), 
which can support biomanufacturing online monitoring and real-time release. 
Correctly quantifying all sources of uncertainty can facilitate optimal learning, guide risk reduction, and support robust control.
\textit{Therefore, the Bayesian KG, accounting for inherent stochasticity and model uncertainty, can be created and used to support integrated bioprocess risk, sensitivity, and predictive analyses.}

Given finite real-world data of the partially observed bioprocess trajectory with size $m$, denoted by $\mathcal{D}_m=\{\pmb\tau_x^{(i)}: i=1,2,\ldots,m\}$ with $\pmb\tau_x\equiv (\pmb{x}_1,\pmb{a}_1,\ldots,\pmb{x}_H,\pmb{a}_H,\pmb{x}_{H+1})$, the model uncertainty is quantified by a posterior distribution,
\begin{equation}
p(\pmb{\theta}|\mathcal{D}_m) \propto p(\pmb{\theta})p(\mathcal{D}_m|\pmb{\theta})=p(\pmb{\theta})\prod_{i=1}^m p\left(\left.\pmb{\tau}_x^{(i)} \right|\pmb{\theta} \right)
\label{eq.posterior}
\nonumber 
\end{equation}
where $p(\pmb{\theta})$ represents the prior distribution. Since the likelihood evaluation of the KG hybrid model, with high fidelity to capture the critical properties of bioprocessing, is intractable, i.e., $p(\pmb{\tau}_x| \pmb\theta)= \int \cdots \int p(\pmb\tau|\pmb\theta) d \pmb{z}_1 \cdots d \pmb{z}_{H+1}$, approximate Bayesian computation sampling with Sequential Monte Carlo (ABC-SMC) is developed to approximate the posterior distribution \cite{zheng2022opportunities,xie2022dynamic}.
In the naive ABC implementation, we draw a candidate sample from the prior $\pmb{\theta} \sim p(\pmb{\theta})$ and then generate a simulation dataset $\mathcal{D}^\star$ from the hybrid model. 
If the simulated dataset $\mathcal{D}^\star$ is ``close" to the observed real-world observations $\mathcal{D}_m$, we accept the sample $\pmb{\theta}$; otherwise reject it.
The accept rate can be very low since it is very computationally expensive to match random trajectories from complex bioprocesses especially under the situations with high stochastic and model uncertainties.
\end{sloppypar}

The ABC-sequential Monte Carlo (ABC-SMC) methods \cite{toni2009approximate,martin2019auxiliary} can improve the sampling efficiency through: (1) generating candidate samples from updated posterior approximates by using sequential importance sampling (SIS); and (2) matching ``well-designed" summary statistics, denoted by ${\eta}(\mathcal{D})$, instead trajectory observations. 
Following the spirit of the auxiliary likelihood-based ABC \cite{martin2019auxiliary}, 
{we create a linear Gaussian dynamic Bayesian network (LG-DBN) auxiliary model and derive  
summary statistics for ABC-SMC that can accelerate online Bayesian inference on KG hybrid models 
\cite{xie2022dynamic}.} 
This simple LG-DBN auxiliary model, in conjunction with SIS, can capture critical biophysicochemical interactions and variations of bioprocess trajectory, ensure the computational efficiency, and enable high quality of inference. \textit{Therefore, the proposed LG-DBN auxiliary likelihood-based ABC-SMC approach can support process soft sensor monitoring, facilitate mechanism online learning, and guide robust process control.}


\vspace{0.1in}

\textbf{Interpretable prediction and sensitivity analysis.}
Given process model parameters and policy parameters, denoted by $(\pmb{\theta},\pmb{\phi})$, the {spatial-temporal interdependencies} of the bioprocess SDP trajectory $\pmb\tau=(\pmb{s}_1,\pmb{a}_1,\ldots,\pmb{s}_H,\pmb{a}_H,\pmb{s}_{H+1})$ is quantified by the joint distribution,
$p(\pmb\tau|\pmb{\theta},\pmb{\phi})=p(\pmb{s}_1)\prod^{H}_{t=1}p(\pmb{s}_{t+1}|\pmb{s}_t,\pmb{a}_t;\pmb{\theta})\pi_{\pmb\phi}(\pmb{a}_t|\pmb{s}_t)$,
which depends on underlying bioprocess mechanisms, sensor network design, and data collection strategies. 
For each batch of production, given inputs denoted by $\mathbf{X}$ (e.g., mRNA sequence, lipid design), we can predict any intermediate or final outputs, denoted by $\mathbf{Y}$ (e.g., CQAs of mRNA-LNP) by using Bayesian KG. The \textit{prediction risk} can be quantified by the posterior predictive distribution, 
$P(\mathbf{Y}| \mathbf{X}) = \int P(\mathbf{Y}| \mathbf{X}, \pmb\theta) p(\pmb{\theta}|\mathcal{D}_m) d \pmb{\theta}$, accounting for both stochastic and model uncertainties.

\textit{We create a Shapley value (SV)-based  sensitivity analysis scheme on the Bayesian KG, called ``KG-SV", to support backward root cause analysis and forward interpretable predictive analysis \cite{xie2022interpretable,zheng2021policy} .}
Since the proposed Bayesian KG-SV can faithfully account for bioprocess causal interdependencies and biophysicochemical interactions, it can correctly assess the effect from each input variation (such as RNA virus mutation), as well as the impact of each source of stochastic and model uncertainties on the prediction risks.
{The criticality assessment of input factors is
based on the estimated values and estimation uncertainties of interpretable KG hybrid model parameters -- such as mechanism pathways in the bioprocess KG from inputs to output} (i.e., 
biomolecular reaction rates, mRNA lipid nanoparticle clustering kinetic parameters).
 Since model uncertainty can be efficiently reduced by most ``informative" data collection and SDP inherent stochasticity can be controlled by decision making, the Bayesian KG based risk, sensitivity, and predictive analyses can identify bottlenecks, guide optimal learning based data collection, and enhance biomanufacturing process CPPs/CQAs specifications
 for QbD.

\subsection{Reinforcement Learning for Bioprocess Design and Control}
\label{sec:RL}

\textit{The proposed Bayesian KG built in conjunction with reinforcement learning (RL) can support long-term prediction and guide interpretable, robust, and optimal decision making.}
Given 
any feasible policy specified by parameters $\pmb{\phi} \in \mathbb{C}$, i.e., $\pmb{a}_t=\pi_t (\pmb{s}_t|\pmb{\phi})$, the optimization of the policy $\pi$ is to maximize the expected accumulated reward,
\begin{equation}
    J(\pmb{\phi}) \equiv \mbox{E}_{\pmb{\theta}\sim p\left(\pmb{\theta}|\mathcal{D}\right)} \left[ \mbox{E}_{\pmb\tau \sim  p(\pmb\tau|{\pmb\theta})}\left[\left.\sum_{t=1}^{H+1} r_t(\pmb{s}_t,\pmb{a}_t)
\right|\pmb{\pi},
{\pmb{{\theta}}}\right] \right],
\label{eq: objective-simple}
\end{equation}
accounting for bioprocess inherent stochasticity and model uncertainty, where $\mathbb{C}$ is a feasible region.
At any $k$-th iteration, we can use the policy gradient to solve the optimization,
\begin{equation}\label{eq:sa}
\pmb{\phi}_{k+1} = \Pi_{\mathbb{C}}\left(\pmb{\phi}_k + \eta_k \nabla{J}\left(\pmb{\phi}_k\right)\right),
\end{equation}
where $\eta_k$ is a suitable 
stepsize and $\Pi_{\mathbb{C}}$ is a projection onto $\mathbb{C}$. 

\vspace{0.1in}
\textbf{Reinforcement learning scheme on Bayesian KG.}
We propose model-based RL scheme on the Bayesian KG \cite{zheng2021policy}, which can provide an insightful prediction on 
how the effect of input factor propagates through bioprocess mechanism pathways and impacts on the outputs.
It can find control policies that are interpretable and robust against heterogeneous model uncertainty, and overcome the key challenges of biopharmaceutical manufacturing, i.e., high complexity, high uncertainty, and very limited process data.
To support real-time control for complex biomanufacturing processes, we provide a provably convergent
stochastic policy gradient optimization and it is computationally efficient through reusing computations associated with similar input-output mechanism pathways.

\vspace{0.1in}
\textbf{Hybrid Model Likelihood Ratio based Historical Observation Reuse.}
Since each experiment run is very computationally expensive especially for multi-scale bioprocess hybrid model, we propose KG assisted multiple important sampling (``KG-MIS") to accelerate policy gradient optimization \cite{zheng2021green}.
Basically, 
we can select and reuse the ``most relevant" historical trajectories, improve policy gradient estimation, and accelerate the search for the optimal robust policy.
For high dimensional SDP,
this study can selectively reuse historical trajectories
having similar underlying distributions with that of target SDP and improve the estimation of policy gradient. 

\begin{sloppypar}
 In classical policy gradient (PG) approach, at any $k$-th iteration,
 the sample average approximation (SAA) is used to estimate the gradient based on $n$ new trajectories generated,
 $\nabla\widehat{J}^{PG}(\pmb\theta_k)=\frac{1}{n}\sum^{n}_{j=1}g\left(\pmb\tau^{(k,j)}|\pmb{\theta}^{(k,j)},\pmb{\phi}_k \right)$ with the scenario gradient $
     g(\pmb\tau|\pmb\theta,\pmb{\phi})=\nabla_{\pmb\theta}\mbox{E}_{\pmb\tau}\left[\left.\sum_{t=1}^H r_t(\pmb{s}_t,\pmb{a}_t)
\right|\pmb\theta,\pmb{\phi}\right]$,
where $\pmb{\theta}^{(k,j)} \sim p(\pmb{\theta}|\mathcal{D}_{m})$ and $\pmb\tau^{(k,j)} \sim p(\pmb\tau|\pmb{\theta}^{(k,j)},\pmb\phi_k)$.
The target SDP mixture distribution 
$p_k(\pmb\tau) = \frac{1}{n} \sum_{j=1}^n
p(\pmb\tau|\pmb{\theta}^{(k,j)},\pmb\phi_k)
$ accounts for both process stochastic and model uncertainties.
Motivated by the studies on multiple important sampling (MIS) \cite{Dong2018,FengGreenSim2017}, 
we create a KG-MIS policy gradient unbiased estimator, 
\begin{equation}
\nabla\widehat{J}^{MIS}(\pmb\phi_k)=\frac{1}{n|U_k|}\sum_{i\in U_k}\sum^n_{j=1}f_{k}\left(\pmb\tau^{(i,j)}\right)g\left.\left(\pmb\tau^{(i,j)}\right|
\pmb{\theta}^{(k,j)}, \pmb{\phi}_k\right)
~~~\mbox{with}
~~~ f_{k}(\pmb\tau) = \frac{
p_k(\pmb\tau)
}
{\sum_{i \in U_k} p(\pmb\tau|\pmb{\theta}_i,\pmb{\phi}_i)/|U_k|}. 
\end{equation}
Since an inappropriate selection of reuse set $U_k$ can lead to the inflated estimation variance of policy gradient, we propose a variance reduction based experience replay criteria  \cite{zheng2021green},
which can automatically select the most relevant historical trajectories generated under different decision policies $\pmb{\phi}$ and model parameters $\pmb{\theta}$ from different posterior distributions resulting from online learning and process control.
We prove that the proposed approach is asymptotically convergent and show it significantly outperforms classical policy gradient approach. Furthermore, 
we extend the proposed KG-MIS framework so that it can select and reuse the most relevant partial trajectories from historical observations \cite{ZhengXie-partialReuse-2022}, i.e., the reuse unit is defined based on state-action transition $(\pmb{s},\pmb{a},\pmb{s}^\prime)$. This study can allow us to flexibly integrate and leverage the relevant information collected from different production lines and facilitate personalized bio-drug manufacturing.

\end{sloppypar}


\section{CHALLENGES AND OPPORTUNITIES FOR OPERATIONS RESEARCH}
\label{sec:summary}

 Increased flexibility, that is necessary to achieve personalized products and manufacturing, should be considered early on as integral part of product design. In fact, for achieving a ``full circle'', not only the manufacturing technology needs to be flexible, but also the drug design and the process control need to support it. Novel operations research approaches and simulation platforms can substantially improve the performance of CMO allowing for larger variability of products with potentially small volume per variant capitalizing upon single use/disposable technologies. 
 
Drug discovery is positively impacted by optimization methods. These should embed scarce data and low fidelity physical models characterizing the existing understanding of bioprocess mechanisms. 
In fact, gray-box search methods are a very active field of research and we believe drug design represents a leading opportunity for further development. The expert-based framework is an example of such approaches, but more efforts are necessary.

Novel simulation methodologies are necessary for analyzing end-to-end biomanufacturing processes and supporting interoperability. Among new upcoming techniques are hybrid modeling, data integration, risk management, and interpretable robust process control. However these are an example and more development in the ares is required. 

\section*{Acknowledgements}
We acknowledge Hua Zheng and Keqi Wang (Northeastern University) for their help on the paper preparation. This work was partially supported by the grants NSF\#2046588-2007861 (PI Pedrielli) and NIIMBL PC4.1-206 (PI Xie).

\bibliographystyle{unsrt}

\bibliography{reference,refgp}

\begin{thebibliography}{10}

\bibitem{Global2020}
Eric Langer.
\newblock {\em Global trends and growth opportunities in the biopharmaceutical
  product development and manufacturing}.
\newblock 2020.
\newblock
  https://www.cphi-online.com/46/resourcefile/10/93/83/CPhI\%20annual\%20rep\%202020\%20vJ.pdf.

\bibitem{Ci15}
R.~Cintron.
\newblock {\em Human Factors Analysis and Classification System Interrater
  Reliability for Biopharmaceutical Manufacturing Investigations}.
\newblock PhD thesis, Walden University, 2015.

\bibitem{Sandle2018epr}
Tim Sandle.
\newblock Strategy for the adoption of single-use technology.
\newblock {\em European Pharmaceutical Review}, 2018.

\bibitem{Blankenship:2020}
Kyle~Blankenship for FiercePharma.
\newblock Samsung scores \$362m deal to help vir scale up covid-19 antibody
  production., 2020 (accessed May 3, 2020).

\bibitem{mock2016automated}
Ulrike Mock, Lauren Nickolay, Brian Philip, Gordon Weng-Kit Cheung, Hong Zhan,
  Ian~CD Johnston, Andrew~D Kaiser, Karl Peggs, Martin Pule, Adrian~J Thrasher,
  et~al.
\newblock Automated manufacturing of chimeric antigen receptor t cells for
  adoptive immunotherapy using clinimacs prodigy.
\newblock {\em Cytotherapy}, 18(8):1002--1011, 2016.

\bibitem{zhu2017closed}
Fenlu Zhu, Nirav Shah, Huiqing Xu, Dina Schneider, Rimas Orentas, Boro
  Dropulic, Parameswaran Hari, and Carolyn~A Keever-Taylor.
\newblock Closed-system manufacturing of cd19 and dual-targeted cd20/19
  chimeric antigen receptor t cells using the clinimacs prodigy device at an
  academic medical center.
\newblock {\em Cytotherapy}, 2017.

\bibitem{lee2019review}
Guesuk Lee, Wongon Kim, Hyunseok Oh, Byeng~D Youn, and Nam~H Kim.
\newblock Review of statistical model calibration and validation—from the
  perspective of uncertainty structures.
\newblock {\em Structural and Multidisciplinary Optimization}, pages 1--26,
  2019.

\bibitem{arendt2012quantification}
Paul~D Arendt, Daniel~W Apley, and Wei Chen.
\newblock Quantification of model uncertainty: Calibration, model discrepancy,
  and identifiability.
\newblock {\em Journal of Mechanical Design}, 134(10), 2012.

\bibitem{wang2019bayesian}
Bo~Wang, Qiong Zhang, and Wei Xie.
\newblock Bayesian sequential data collection for stochastic simulation
  calibration.
\newblock {\em European Journal of Operational Research}, 277(1):300--316,
  2019.

\bibitem{saltelli2008global}
Andrea Saltelli, Marco Ratto, Terry Andres, Francesca Campolongo, Jessica
  Cariboni, Debora Gatelli, Michaela Saisana, and Stefano Tarantola.
\newblock {\em Global sensitivity analysis: the primer}.
\newblock John Wiley \& Sons, 2008.

\bibitem{baroni2014general}
Gabriele Baroni and Stefano Tarantola.
\newblock A general probabilistic framework for uncertainty and global
  sensitivity analysis of deterministic models: A hydrological case study.
\newblock {\em Environmental Modelling \& Software}, 51:26--34, 2014.

\bibitem{lakhdar2007multiobjective}
K~Lakhdar, J~Savery, LG~Papageorgiou, and SS~Farid.
\newblock Multiobjective long-term planning of biopharmaceutical manufacturing
  facilities.
\newblock {\em Biotechnology Progress}, 23(6):1383--1393, 2007.

\bibitem{leachman2014automated}
Robert~C. Leachman, Lenrick Johnston, Shan Li, and Zuo-Jun Shen.
\newblock An automated planning engine for biopharmaceutical production.
\newblock {\em European Journal of Operational Research}, 238(1):327--338,
  2014.

\bibitem{fleischhacker2011planning}
Adam~J Fleischhacker and Yao Zhao.
\newblock Planning for demand failure: A dynamic lot size model for clinical
  trial supply chains.
\newblock {\em European Journal of Operational Research}, 211(3):496--506,
  2011.

\bibitem{kulkarni2015modular}
Niranjan~S. Kulkarni.
\newblock A modular approach for modeling active pharmaceutical ingredient
  manufacturing plant: A case study.
\newblock In {\em Proceedings of the 2015 Winter Simulation Conference}, pages
  2260--2271, Piscataway, New Jersey, 2015. Institute of Electrical and
  Electronics Engineers, Inc.

\bibitem{martagan2018performance}
Tugce Martagan, Ananth Krishnamurthy, Peter~A Leland, and Christos~T
  Maravelias.
\newblock Performance guarantees and optimal purification decisions for
  engineered proteins.
\newblock {\em Operations Research}, 66(1):18--41, 2018.

\bibitem{green2014toehold}
Alexander~A Green, Pamela~A Silver, James~J Collins, and Peng Yin.
\newblock Toehold switches: de-novo-designed regulators of gene expression.
\newblock {\em Cell}, 159(4):925--939, 2014.

\bibitem{geary2014single}
Cody Geary, Paul~WK Rothemund, and Ebbe~S Andersen.
\newblock A single-stranded architecture for cotranscriptional folding of {RNA}
  nanostructures.
\newblock {\em Science}, 345(6198):799--804, 2014.

\bibitem{han2017single}
Dongran Han, Xiaodong Qi, Cameron Myhrvold, Bei Wang, Mingjie Dai, Shuoxing
  Jiang, Maxwell Bates, Yan Liu, Byoungkwon An, Fei Zhang, et~al.
\newblock Single-stranded {DNA} and {RNA} origami.
\newblock {\em Science}, 358(6369):eaao2648, 2017.

\bibitem{liu2022expertrna}
Menghan Liu, Erik Poppleton, Giulia Pedrielli, Petr {\v{S}}ulc, and Dimitri~P
  Bertsekas.
\newblock Expertrna: A new framework for rna secondary structure prediction.
\newblock {\em INFORMS Journal on Computing}, 2022.

\bibitem{carell2012structure}
Thomas Carell, Caterina Brandmayr, Antje Hienzsch, Markus M{\"u}ller, David
  Pearson, Veronika Reiter, Ines Thoma, Peter Thumbs, and Mirko Wagner.
\newblock Structure and function of noncanonical nucleobases.
\newblock {\em Angewandte Chemie International Edition}, 51(29):7110--7131,
  2012.

\bibitem{calonaci2020machine}
Nicola Calonaci, Alisha Jones, Francesca Cuturello, Michael Sattler, and
  Giovanni Bussi.
\newblock Machine learning a model for {RNA} structure prediction.
\newblock {\em arXiv preprint arXiv:2004.00351}, 2020.

\bibitem{cruz2012rna}
Jos{\'e}~Almeida Cruz, Marc-Fr{\'e}d{\'e}rick Blanchet, Michal Boniecki,
  Janusz~M Bujnicki, Shi-Jie Chen, Song Cao, Rhiju Das, Feng Ding, Nikolay~V
  Dokholyan, Samuel~Coulbourn Flores, et~al.
\newblock {RNA}-puzzles: a casp-like evaluation of {RNA} three-dimensional
  structure prediction.
\newblock {\em Rna}, 18(4):610--625, 2012.

\bibitem{reuter2010rnastructure}
Jessica~S Reuter and David~H Mathews.
\newblock {RNA}structure: software for {RNA} secondary structure prediction and
  analysis.
\newblock {\em BMC bioinformatics}, 11(1):129, 2010.

\bibitem{hofacker2003vienna}
Ivo~L Hofacker.
\newblock Vienna {RNA} secondary structure server.
\newblock {\em Nucleic acids research}, 31(13):3429--3431, 2003.

\bibitem{zadeh2011nupack}
Joseph~N Zadeh, Conrad~D Steenberg, Justin~S Bois, Brian~R Wolfe, Marshall~B
  Pierce, Asif~R Khan, Robert~M Dirks, and Niles~A Pierce.
\newblock Nupack: analysis and design of nucleic acid systems.
\newblock {\em Journal of computational chemistry}, 32(1):170--173, 2011.

\bibitem{zuker1981optimal}
Michael Zuker and Patrick Stiegler.
\newblock Optimal computer folding of large {RNA} sequences using
  thermodynamics and auxiliary information.
\newblock {\em Nucleic acids research}, 9(1):133--148, 1981.

\bibitem{mathews1999expanded}
David~H Mathews, Jeffrey Sabina, Michael Zuker, and Douglas~H Turner.
\newblock Expanded sequence dependence of thermodynamic parameters improves
  prediction of {RNA} secondary structure.
\newblock {\em Journal of molecular biology}, 288(5):911--940, 1999.

\bibitem{xia1998thermodynamic}
Tianbing Xia, John SantaLucia~Jr, Mark~E Burkard, Ryszard Kierzek, Susan~J
  Schroeder, Xiaoqi Jiao, Christopher Cox, and Douglas~H Turner.
\newblock Thermodynamic parameters for an expanded nearest-neighbor model for
  formation of rna duplexes with watson- crick base pairs.
\newblock {\em Biochemistry}, 37(42):14719--14735, 1998.

\bibitem{andronescu2010computational}
Mirela Andronescu, Anne Condon, Holger~H Hoos, David~H Mathews, and Kevin~P
  Murphy.
\newblock Computational approaches for {RNA} energy parameter estimation.
\newblock {\em RNA}, 16(12):2304--2318, 2010.

\bibitem{angela2018computationally}
M~Yu Angela, Paul~M Gasper, Eric~J Strobel, Kyle~E Watters, Alan~A Chen, and
  Julius~B Lucks.
\newblock Computationally reconstructing cotranscriptional {RNA} folding
  pathways from experimental data reveals rearrangement of non-native folding
  intermediates.
\newblock {\em bioRxiv}, page 379222, 2018.

\bibitem{isambert2000modeling}
Herv{\'e} Isambert and Eric~D Siggia.
\newblock Modeling {RNA} folding paths with pseudoknots: application to
  hepatitis delta virus ribozyme.
\newblock {\em PNAS}, 97(12):6515--6520, 2000.

\bibitem{sun2018predicting}
Ting-ting Sun, Chenhan Zhao, and Shi-Jie Chen.
\newblock Predicting cotranscriptional folding kinetics for riboswitch.
\newblock {\em The Journal of Physical Chemistry B}, 122(30):7484--7496, 2018.

\bibitem{do2006contrafold}
Chuong~B Do, Daniel~A Woods, and Serafim Batzoglou.
\newblock Contrafold: {RNA} secondary structure prediction without
  physics-based models.
\newblock {\em Bioinformatics}, 22(14):e90--e98, 2006.

\bibitem{lucks2011multiplexed}
Julius~B Lucks, Stefanie~A Mortimer, Cole Trapnell, Shujun Luo, Sharon Aviran,
  Gary~P Schroth, Lior Pachter, Jennifer~A Doudna, and Adam~P Arkin.
\newblock Multiplexed {RNA} structure characterization with selective
  2'-hydroxyl acylation analyzed by primer extension sequencing (shape-seq).
\newblock {\em Proceedings of the National Academy of Sciences},
  108(27):11063--11068, 2011.

\bibitem{low2010shape}
Justin~T Low and Kevin~M Weeks.
\newblock Shape-directed {RNA} secondary structure prediction.
\newblock {\em Methods}, 52(2):150--158, 2010.

\bibitem{aghaeepour2013ensemble}
Nima Aghaeepour and Holger~H Hoos.
\newblock Ensemble-based prediction of {RNA} secondary structures.
\newblock {\em BMC bioinformatics}, 14(1):139, 2013.

\bibitem{wayment2020rna}
Hannah~K Wayment-Steele, Wipapat Kladwang, Eterna Participants, and Rhiju Das.
\newblock {RNA} secondary structure packages ranked and improved by
  high-throughput experiments.
\newblock {\em BioRxiv}, 2020.

\bibitem{miao2017rna}
Zhichao Miao, Ryszard~W Adamiak, Maciej Antczak, Robert~T Batey, Alexander~J
  Becka, Marcin Biesiada, Micha{\l}~J Boniecki, Janusz~M Bujnicki, Shi-Jie
  Chen, Clarence~Yu Cheng, et~al.
\newblock {RNA}-puzzles round iii: 3d {RNA} structure prediction of five
  riboswitches and one ribozyme.
\newblock {\em {Rna}}, 23(5):655--672, 2017.

\bibitem{watkins2020farfar2}
Andrew~Martin Watkins, Ramya Rangan, and Rhiju Das.
\newblock Farfar2: Improved de novo rosetta prediction of complex global {RNA}
  folds.
\newblock {\em Structure}, 2020.

\bibitem{thevenet2012pep}
Pierre Th{\'e}venet, Yimin Shen, Julien Maupetit, Fr{\'e}d{\'e}ric Guyon,
  Philippe Derreumaux, and Pierre Tuff{\'e}ry.
\newblock Pep-fold: an updated de novo structure prediction server for both
  linear and disulfide bonded cyclic peptides.
\newblock {\em Nucleic acids research}, 40(W1):W288--W293, 2012.

\bibitem{shen2014improved}
Yimin Shen, Julien Maupetit, Philippe Derreumaux, and Pierre Tufféry.
\newblock Improved pep-fold approach for peptide and miniprotein structure
  prediction.
\newblock {\em Journal of chemical theory and computation}, 10(10):4745--4758,
  2014.

\bibitem{lamiable2016pep}
Alexis Lamiable, Pierre Th{\'e}venet, Julien Rey, Marek Vavrusa, Philippe
  Derreumaux, and Pierre Tuff{\'e}ry.
\newblock Pep-fold3: faster de novo structure prediction for linear peptides in
  solution and in complex.
\newblock {\em Nucleic acids research}, 44(W1):W449--W454, 2016.

\bibitem{jin2020awsem}
Shikai Jin, Vinicius~G Contessoto, Mingchen Chen, Nicholas~P Schafer, Wei Lu,
  Xun Chen, Carlos Bueno, Arya Hajitaheri, Brian~J Sirovetz, Aram Davtyan,
  et~al.
\newblock Awsem-suite: a protein structure prediction server based on
  template-guided, coevolutionary-enhanced optimized folding landscapes.
\newblock {\em Nucleic acids research}, 48(W1):W25--W30, 2020.

\bibitem{chaudhury2010pyrosetta}
Sidhartha Chaudhury, Sergey Lyskov, and Jeffrey~J Gray.
\newblock Pyrosetta: a script-based interface for implementing molecular
  modeling algorithms using rosetta.
\newblock {\em Bioinformatics}, 26(5):689--691, 2010.

\bibitem{alquraishi2019alphafold}
Mohammed AlQuraishi.
\newblock Alphafold at casp13.
\newblock {\em Bioinformatics}, 35(22):4862--4865, 2019.

\bibitem{marx2022method}
Vivien Marx.
\newblock Method of the year: protein structure prediction.
\newblock {\em Nature methods}, 19(1):5--10, 2022.

\bibitem{sellami2021virtual}
Asma Sellami, Manon R{\'e}au, Florent Langenfeld, Nathalie Lagarde, and
  Matthieu Montes.
\newblock Virtual libraries for docking methods: Guidelines for the selection
  and the preparation.
\newblock In {\em Molecular Docking for Computer-Aided Drug Design}, pages
  99--117. Elsevier, 2021.

\bibitem{stanzione2021use}
Francesca Stanzione, Ilenia Giangreco, and Jason~C Cole.
\newblock Use of molecular docking computational tools in drug discovery.
\newblock {\em Progress in Medicinal Chemistry}, 60:273--343, 2021.

\bibitem{friesner2004glide}
Richard~A Friesner, Jay~L Banks, Robert~B Murphy, Thomas~A Halgren, Jasna~J
  Klicic, Daniel~T Mainz, Matthew~P Repasky, Eric~H Knoll, Mee Shelley, Jason~K
  Perry, et~al.
\newblock Glide: a new approach for rapid, accurate docking and scoring. 1.
  method and assessment of docking accuracy.
\newblock {\em Journal of medicinal chemistry}, 47(7):1739--1749, 2004.

\bibitem{halgren2004glide}
Thomas~A Halgren, Robert~B Murphy, Richard~A Friesner, Hege~S Beard, Leah~L
  Frye, W~Thomas Pollard, and Jay~L Banks.
\newblock Glide: a new approach for rapid, accurate docking and scoring. 2.
  enrichment factors in database screening.
\newblock {\em Journal of medicinal chemistry}, 47(7):1750--1759, 2004.

\bibitem{rarey1996fast}
Matthias Rarey, Bernd Kramer, Thomas Lengauer, and Gerhard Klebe.
\newblock A fast flexible docking method using an incremental construction
  algorithm.
\newblock {\em Journal of molecular biology}, 261(3):470--489, 1996.

\bibitem{jones1997development}
Gareth Jones, Peter Willett, Robert~C Glen, Andrew~R Leach, and Robin Taylor.
\newblock Development and validation of a genetic algorithm for flexible
  docking.
\newblock {\em Journal of molecular biology}, 267(3):727--748, 1997.

\bibitem{hart1992multiple}
Trevor~N Hart and Randy~J Read.
\newblock A multiple-start monte carlo docking method.
\newblock {\em Proteins: Structure, Function, and Bioinformatics},
  13(3):206--222, 1992.

\bibitem{korb2006plants}
Oliver Korb, Thomas St{\"u}tzle, and Thomas~E Exner.
\newblock Plants: Application of ant colony optimization to structure-based
  drug design.
\newblock In {\em International workshop on ant colony optimization and swarm
  intelligence}, pages 247--258. Springer, 2006.

\bibitem{pei2006psi}
Jianfeng Pei, Qi~Wang, Zhenming Liu, Qingliang Li, Kun Yang, and Luhua Lai.
\newblock Psi-dock: Towards highly efficient and accurate flexible ligand
  docking.
\newblock {\em Proteins: Structure, Function, and Bioinformatics},
  62(4):934--946, 2006.

\bibitem{huang2006benchmarking}
Niu Huang, Brian~K Shoichet, and John~J Irwin.
\newblock Benchmarking sets for molecular docking.
\newblock {\em Journal of medicinal chemistry}, 49(23):6789--6801, 2006.

\bibitem{lagarde2015benchmarking}
Nathalie Lagarde, Jean-Fran{\c{c}}ois Zagury, and Matthieu Montes.
\newblock Benchmarking data sets for the evaluation of virtual ligand screening
  methods: review and perspectives.
\newblock {\em Journal of chemical information and modeling}, 55(7):1297--1307,
  2015.

\bibitem{mysinger2012directory}
Michael~M Mysinger, Michael Carchia, John~J Irwin, and Brian~K Shoichet.
\newblock Directory of useful decoys, enhanced (dud-e): better ligands and
  decoys for better benchmarking.
\newblock {\em Journal of medicinal chemistry}, 55(14):6582--6594, 2012.

\bibitem{sterling2015zinc}
Teague Sterling and John~J Irwin.
\newblock Zinc 15--ligand discovery for everyone.
\newblock {\em Journal of chemical information and modeling},
  55(11):2324--2337, 2015.

\bibitem{kim2019pubchem}
Sunghwan Kim, Jie Chen, Tiejun Cheng, Asta Gindulyte, Jia He, Siqian He,
  Qingliang Li, Benjamin~A Shoemaker, Paul~A Thiessen, Bo~Yu, et~al.
\newblock Pubchem 2019 update: improved access to chemical data.
\newblock {\em Nucleic acids research}, 47(D1):D1102--D1109, 2019.

\bibitem{newman2020natural}
David~J Newman and Gordon~M Cragg.
\newblock Natural products as sources of new drugs over the nearly four decades
  from 01/1981 to 09/2019.
\newblock {\em Journal of natural products}, 83(3):770--803, 2020.

\bibitem{morrison2013natural}
Karen~C Morrison and Paul~J Hergenrother.
\newblock Natural products as starting points for the synthesis of complex and
  diverse compounds.
\newblock {\em Natural product reports}, 31(1):6--14, 2013.

\bibitem{rodrigues2016counting}
Tiago Rodrigues, Daniel Reker, Petra Schneider, and Gisbert Schneider.
\newblock Counting on natural products for drug design.
\newblock {\em Nature chemistry}, 8(6):531--541, 2016.

\bibitem{sorokina2020review}
Maria Sorokina and Christoph Steinbeck.
\newblock Review on natural products databases: Where to find data in 2020.
\newblock {\em Journal of Cheminformatics}, 12(1):1--51, 2020.

\bibitem{chen2017data}
Ya~Chen, Christina de~Bruyn~Kops, and Johannes Kirchmair.
\newblock Data resources for the computer-guided discovery of bioactive natural
  products.
\newblock {\em Journal of chemical information and modeling}, 57(9):2099--2111,
  2017.

\bibitem{huang2010scoring}
Sheng-You Huang, Sam~Z Grinter, and Xiaoqin Zou.
\newblock Scoring functions and their evaluation methods for protein--ligand
  docking: recent advances and future directions.
\newblock {\em Physical Chemistry Chemical Physics}, 12(40):12899--12908, 2010.

\bibitem{ain2015machine}
Qurrat~Ul Ain, Antoniya Aleksandrova, Florian~D Roessler, and Pedro~J
  Ballester.
\newblock Machine-learning scoring functions to improve structure-based binding
  affinity prediction and virtual screening.
\newblock {\em Wiley Interdisciplinary Reviews: Computational Molecular
  Science}, 5(6):405--424, 2015.

\bibitem{huang2006molecular}
Niu Huang, Chakrapani Kalyanaraman, Katarzyna Bernacki, and Matthew~P Jacobson.
\newblock Molecular mechanics methods for predicting protein--ligand binding.
\newblock {\em Physical Chemistry Chemical Physics}, 8(44):5166--5177, 2006.

\bibitem{gohlke2000knowledge}
Holger Gohlke, Manfred Hendlich, and Gerhard Klebe.
\newblock Knowledge-based scoring function to predict protein-ligand
  interactions.
\newblock {\em Journal of molecular biology}, 295(2):337--356, 2000.

\bibitem{krammer2005ligscore}
Andr{\'e} Krammer, Paul~D Kirchhoff, X~Jiang, CM~Venkatachalam, and Marvin
  Waldman.
\newblock Ligscore: a novel scoring function for predicting binding affinities.
\newblock {\em Journal of Molecular Graphics and Modelling}, 23(5):395--407,
  2005.

\bibitem{guvench2009computational}
Olgun Guvench and Alexander~D MacKerell~Jr.
\newblock Computational evaluation of protein--small molecule binding.
\newblock {\em Current opinion in structural biology}, 19(1):56--61, 2009.

\bibitem{ballester2010machine}
Pedro~J Ballester and John~BO Mitchell.
\newblock A machine learning approach to predicting protein--ligand binding
  affinity with applications to molecular docking.
\newblock {\em Bioinformatics}, 26(9):1169--1175, 2010.

\bibitem{zheng2022opportunities}
Hua Zheng, Wei Xie, Keqi Wang, and Zheng Li.
\newblock Opportunities of hybrid model-based reinforcement learning for cell
  therapy manufacturing process development and control.
\newblock {\em arXiv preprint arXiv:2201.03116}, 2022.

\bibitem{xie2022interpretable}
Wei Xie, Bo~Wang, Cheng Li, Dongming Xie, and Jared Auclair.
\newblock Interpretable biomanufacturing process risk and sensitivity analyses
  for quality-by-design and stability control.
\newblock {\em Naval Research Logistics (NRL)}, 69(3):461--483, 2022.

\bibitem{patro2002protein}
Sugunakar~Y Patro, Erwin Freund, and Byeong~S Chang.
\newblock Protein formulation and fill-finish operations.
\newblock 2002.

\bibitem{guideline2009pharmaceutical}
ICH Harmonised~Tripartite Guideline et~al.
\newblock Pharmaceutical development.
\newblock {\em Q8 (R2) Current Step}, 4, 2009.

\bibitem{buschmann2021nanomaterial}
Michael~D Buschmann, Manuel~J Carrasco, Suman Alishetty, Mikell Paige,
  Mohamad~Gabriel Alameh, and Drew Weissman.
\newblock Nanomaterial delivery systems for mrna vaccines.
\newblock {\em Vaccines}, 9(1):65, 2021.

\bibitem{xie2022dynamic}
Wei Xie, Keqi Wang, Hua Zheng, and Ben Feng.
\newblock Dynamic bayesian network auxiliary abc-smc for hybrid model bayesian
  inference to accelerate biomanufacturing process mechanism learning and
  robust control.
\newblock {\em arXiv preprint arXiv:2205.02410}, 2022.

\bibitem{toni2009approximate}
Tina Toni, David Welch, Natalja Strelkowa, Andreas Ipsen, and Michael~PH
  Stumpf.
\newblock Approximate bayesian computation scheme for parameter inference and
  model selection in dynamical systems.
\newblock {\em Journal of the Royal Society Interface}, 6(31):187--202, 2009.

\bibitem{martin2019auxiliary}
Gael~M Martin, Brendan~PM McCabe, David~T Frazier, Worapree Maneesoonthorn, and
  Christian~P Robert.
\newblock Auxiliary likelihood-based approximate bayesian computation in state
  space models.
\newblock {\em Journal of Computational and Graphical Statistics},
  28(3):508--522, 2019.

\bibitem{zheng2021policy}
Hua Zheng, Wei Xie, Ilya~O Ryzhov, and Dongming Xie.
\newblock Policy optimization in bayesian network hybrid models of
  biomanufacturing processes.
\newblock {\em arXiv preprint arXiv:2105.06543}, 2021.

\bibitem{zheng2021green}
Hua Zheng, Wei Xie, and M~Ben Feng.
\newblock Green simulation assisted policy gradient to accelerate stochastic
  process control.
\newblock {\em arXiv preprint arXiv:2110.08902}, 2021.

\bibitem{Dong2018}
J.~{Dong}, M.~B. {Feng}, and B.~L. {Nelson}.
\newblock Unbiased metamodeling via likelihood ratios.
\newblock In {\em 2018 Winter Simulation Conference (WSC)}, pages 1778--1789,
  Dec 2018.

\bibitem{FengGreenSim2017}
Mingbin Feng and Jeremy Staum.
\newblock Green simulation: Reusing the output of repeated experiments.
\newblock {\em ACM Transactions on Modeling and Computer Simulation (TOMACS)},
  27(4):23:1--23:28, October 2017.

\bibitem{ZhengXie-partialReuse-2022}
H.~Zheng and W.~Xie.
\newblock Variance reduction based partial trajectory reuse to accelerate
  policy gradient optimization.
\newblock In {\em Proceedings of the 2022 Winter Simulation Conference},
  Piscataway, New Jersey, 2022. Institute of Electrical and Electronics
  Engineers, Inc.

\end{thebibliography}


\begin{thebibliography}{}

\bibitem[\protect\citeauthoryear{Banks, Carson, Nelson, and Nicol}{Banks
  et~al.}{2000}]{bcnn:simulation}
Banks, J., J.~S. Carson, B.~L. Nelson, and D.~M. Nicol. 2000.
\newblock {\em {D}iscrete-{E}vent {S}ystem {S}imulation\/}. 3rd ed.
\newblock Upper Saddle River, New Jersey: Prentice-Hall, Inc.


\bibitem[\protect\citeauthoryear{Cheng}{Cheng}{1994}]{cheng:input94}
Cheng, R. C.~H. 1994.
\newblock ``Selecting Input Models''.
\newblock In {\em Proceedings of the 1994 Winter Simulation Conference}, edited
  by\ J.~D. Tew, S.~Manivannan, D.~A. Sadowski, and A.~F. Seila,  184--191.
\newblock Piscataway, New Jersey: Institute of Electrical and Electronics
  Engineers, Inc.

\bibitem[\protect\citeauthoryear{Chien}{Chien}{1989}]{chi89}
Chien, C. 1989.
\newblock ``Small Sample Theory for Steady State Confidence Intervals''.
\newblock Technical Report No. 37, Department of Operations Research, Stanford
  University, Stanford, California.

\bibitem[\protect\citeauthoryear{Gupta, Nagel, and Panchapakesan}{Gupta
  et~al.}{1973}]{gupta:mnormal}
Gupta, S.~S., K.~Nagel, and S.~Panchapakesan. 1973.
\newblock ``On the Order Statistics from Equally Correlated Normal Random
  Variables''.
\newblock {\em Biometrika\/}~60(2):403--413.


\bibitem[\protect\citeauthoryear{Hammersley and Handscomb}{Hammersley and
  Handscomb}{1964}]{hammersley:montecarlo}
Hammersley, J.~M., and D.~C. Handscomb. 1964.
\newblock {\em Monte Carlo Methods}.
\newblock London: Methuen.


\bibitem[\protect\citeauthoryear{Law and Kelton}{Law and
  Kelton}{2000}]{law:simulationc}
Law, A.~M., and W.~D. Kelton. 2000.
\newblock {\em Simulation Modeling \& Analysis\/}. 3rd ed.
\newblock New York: McGraw-Hill, Inc.


\bibitem[\protect\citeauthoryear{Powell and Mustafee}{Powell and
  Mustafee}{2017}]{powell2017widening}
Powell, J.~H., and N.~Mustafee. 2017.
\newblock ``Widening requirements capture with soft methods: an investigation
  of hybrid M\&S studies in health care''.
\newblock {\em Journal of the Operational Research
  Society\/}~68(10):1211--1222.


\bibitem[\protect\citeauthoryear{Rabe, Dross, and Wuttke}{Rabe
  et~al.}{2017}]{rabe:combining}
Rabe, M., F.~Dross, and A.~Wuttke. 2017.
\newblock ``Combining a Discrete-event Simulation Model of a Logistics Network
  with Deep Reinforcement Learning''.
\newblock In {\em Proceedings of the MIC and MAEB 2017 Conferences}.
\newblock July 4\textsuperscript{th}-7\textsuperscript{th}, Barcelona, Spain,
  765-774.

\bibitem[\protect\citeauthoryear{Schruben}{Schruben}{1979}]{sch79}
Schruben, L.~W. 1979.
\newblock ``Designing Correlation Induction Strategies for Simulation
  Experiments''.
\newblock In {\em Current Issues in Computer Simulation}, edited by\ N.~R. Adam
  and A.~Dogramaci,  235--256. New York: Academic Press.

\bibitem[\protect\citeauthoryear{Steiger}{Steiger}{1999}]{ste99}
Steiger, N.~M. 1999.
\newblock {\em Improved Batching for Confidence Interval Construction in
  Steady-State Simulation}.
\newblock {Ph.D.} thesis, Department of Industrial Engineering, North Carolina
  State University, Raleigh, North Carolina.
\newblock \url{http://www.lib.ncsu.edu/resolver/1840.16/4713}.

\bibitem[\protect\citeauthoryear{{The University of Chicago Press}}{{The
  University of Chicago Press}}{2010}]{chicago03}
{The University of Chicago Press} 2010.
\newblock {\em The Chicago Manual of Style\/}. 16th ed.
\newblock Chicago: The University of Chicago Press.
\newblock \url{http://www.chicagomanualofstyle.org}.

\bibitem[\protect\citeauthoryear{{WSC}}{{WSC}}{2022}]{WSC}
{WSC} 2022.
\newblock ``Winter Simulation Conference''.
\newblock accessed 12\textsuperscript{nd} January 2022.

\end{thebibliography}

\section*{AUTHOR BIOGRAPHIES}

\noindent {\bf WEI XIE} is an assistant professor in MIE at Northeastern University.
Her research
interests include interpretable Artificial Intelligence (AI), machine learning (ML), 
computer simulation, digital twin, data analytics, and stochastic optimization for cyber-physical system risk management, learning, and automation.
Her email address is \href{w.xie@northeastern.edu}{w.xie@northeastern.edu}. Her website is \href{http://www1.coe.neu.edu/~wxie/}{http://www1.coe.neu.edu/$\sim$wxie/}
\vspace{0.05in}

\noindent {\bf GIULIA PEDRIELLI} is Assistant Professor for the School of Computing and Augmented Intelligence at Arizona State University. Her expertise is in the design and analysis of sampling based algorithms, Monte Carlo methods, reinforcement learning, and dynamic simulation. Applications range from molecular design, process design and control, testing and verification of Cyber Physical Systems. Her email address is \href{giulia.pedrielli@asu.edu}{giulia.pedrielli@asu.edu}, her webpage can be found at \href{https://www.gpedriel.com/}{https://www.gpedriel.com/}.\\

\end{document}